\documentclass[10pt. twocolumn,journal]{IEEEtran}
\usepackage{amsmath,amssymb,amsthm,mathrsfs,amsfonts,dsfont,mathtools}
\usepackage[caption=false]{subfig}
\usepackage[ruled,lined,linesnumbered]{algorithm2e}
\usepackage{setspace}
\usepackage{tablefootnote, cite, color}
\usepackage{graphicx,epstopdf}
\usepackage{bmpsize}
\usepackage{hyperref}

\newtheorem{theorem}{Theorem}
\newtheorem{lemma}{Lemma}
\newtheorem{proposition}{Proposition}

\begin{document}
\title{Analysis on $ 60 $ GHz Wireless Communications with Beamwidth-Dependent Misalignment}
\author{Guang~Yang, \emph{Student Member, IEEE}, Jinfeng~Du, \emph{Member, IEEE} and~Ming~Xiao, \emph{Senior Member, IEEE}
%\thanks{This work was supported partly by National Natural Science Foundation of China under Grant 61371105, National 973 Programs 2013CB329001 and EU Marie Curie Project, QUICK, No. 612652.}
\thanks{Guang Yang is with  the Communication Theory Department, Royal Institute of Technology,
Stockholm, Sweden (Email: gy@kth.se).}
\thanks{Jinfeng Du is with Nokia Bell Labs, Holmdel, NJ 07733, United States (Email: jinfeng.du@bell-labs.com).}
\thanks{Ming Xiao is with the Communication Theory Department, Royal Institute of Technology,
Stockholm, Sweden (Email: mingx@kth.se).}
}
\maketitle

\begin{abstract}
High speed wireless access on 60 GHz spectrum relies on high-gain directional antennas  to overcome the severe signal attenuation. However, perfect alignment between transmitting and receiving antenna beams is rare in practice and overheard signals from concurrent transmissions may cause significant interference. 
In this paper we analyze the impact of antenna beam misalignment on the system performance of 60 GHz wireless access.  
We quantify the signal power loss caused by beam misalignment and the interference power accumulated from neighboring concurrent transmissions whose signals are leaked either via the main-beam pointing in the similar direction or via side-lobe emission, and derive the probability distribution of the signal to interference plus noise power ratio (SINR).  
For scenarios where interfering transmitters are distributed uniformly at random, we derive upper and lower bounds on the cumulative distribution function (abbreviated as CDF or c.d.f.) of SINR, which can be easily applied to evaluate system performance. 
We validate our analytical results by simulations where random nodes are uniformly distributed within a circular hall, and  evaluate the sensitivity of average throughput and outage probability against two parameters: the half-power (3 dB) beamwidth to main-lobe beamwidth ratio and the beam misalignment deviation to main-lobe beamwidth ratio.
Our results indicate that the derived lower bound performs well when the half-power beamwidth to main-lobe beamwidth ratio or the number of concurrent transmission links is small. When the number of active links is high,  it is desirable in antenna design to balance the degradation caused by beam misalignment (wider beam is better) and the interference from concurrent transmission (narrower beam is better).
\end{abstract}


\section{Introduction}\label{sec:introduction}

The proliferation of diverse applications and  demands of high speed wireless access \cite{cisco2015visual} drives the rapid development of wireless communication on $ 60 $~GHz band, advocated by many academical and industrial bodies, e.g., IEEE 802.11ad Task Group \cite{802.11ad}, IEEE 802.15.3 Task Group 3c\cite{802.15.3c}, and Wireless Gigabit Alliance (WiGig).
Within the $ 60$~GHz band, the radios encounter many propagation challenges, such as the severe path loss, weak reflection and diffusion, and high penetration loss \cite{moraitis2006measurements, geng2009millimeter}, and therefore the deployment of high-gain directional antennas (arrays) is required. Besides, high directionality has other benefits in systems with concurrent transmissions: it enables high spatial multiplexing to boost the network capacity within a unit area; it lowers the probability of strong interference among current transmissions.  

The benefits of directional antennas and the impact of beam misalignment on the performance of wireless networks have been studied in~\cite{ramanathan2005ad, yu2006performance, epple2006using, wildman2014joint} using simplified beam patterns. 
In general, a narrower beamwidth corresponds to a higher antenna gain and lower probability of experiencing strong interference from concurrent transmissions, which may contribute to significant improvement in network capacity per unit area~\cite{wildman2014joint}.
In most of previous study, the radiation pattern of directional antennas is usually modeled in an idealized fashion, e.g., a constant large antenna gain within the narrow main-lobe and zero else where. This idealized radiation pattern, often referred as the ``flat-top model'', is widely used~\cite{wieselthier2002energy, kang2003power,singh2011interference}  for system level performance analysis.
However, in practice, the radiation patterns of antennas largely depend on their implementation and are usually more more complex: 
  the main-lobe gain is not constant and the side-lobe radiation is non-zero. As the density of nodes increases, the effect of side-lobe radiation and the gradual reduction of main-lobe gain caused by beam misalignment cannot be ignored any more. The maximum beam-forming gain, which can be achieved only if the main-lobe beams of directional transmitting and receiving antennas are perfectly aligned, is rare due to practical implementation constraints. The origin of beam misalignment can be coarsely divided into two categories: imperfection of existing antenna and beamforming techniques~\cite{yu2006performance, shen2005phase, li2006outage}, such as the analog beamforming impairments, array perturbations,  oscillator locking-range based phase error, and the direction-of-arrival (DoA) estimation errors;  mobility of communication terminals \cite{doff2015sensor,hur2013millimeter}, which invokes tracking error and system reaction delay. Therefore, it is crucial to study the beam pattern and beam alignment error and quantify their impacts on performance degradation. 

In recent years, numerous efforts have been devoted in mimicking practical directional antennas and some plausible models are established, e.g., the piece-wised model~\cite{akoum2012coverage, baccelli2009stochastic} and the 3GPP model \cite{3gpp2014}.
The impact of radiation pattern and beam alignment on the performance of directional transmissions has been studied in some recent publications. For instance, in \cite{cai2010rex, qiao2011enabling}, directional antennas considering the side-lobe effect are exploited for mmWave wireless personal area networks (WPAN), and the spatial multiplexing
gain, impact of radiation efficiency and fairness are discussed. Besides, the side-lobe effect has been studied using a piecewise linear model in \cite{akoum2012coverage}. Other related efforts can be seen in \cite{yi2003capacity, wang2010capacity, zhang2010capacity}. %In a recent work \cite{wildman2014joint}, the trade-off between the beam misalignment, beamwidth, and throughput in a directional wireless network is investigated.

In this paper, we adopt a close-to-reality antenna radiation pattern established in the 3GPP standard~\cite{3gpp2014}, where
 the non-constant main-lobe gain and the nonzero side-lobe radiation gain are correlated via a  total radiated power constraint. We  measure the beam misalignment and the half-power ($ 3 $-dB) beamwidth by the ratio between their absolute value and the main-lobe beamwidth, and investigate the effects of radiation pattern and misalignment on performance degradation of $ 60 $~GHz  wireless systems.  We derive the probability distribution of the signal to interference plus noise power ratio (SINR), where the received signal power degrades owing to the imperfection of beam alignment, and the interference power is accumulated through signals leaked from either the side-lobe radiation or the main-lobe beam of surrounding concurrent links.
We also establish upper and lower bounds for the CDF of SINR to facilitate the computation in characterizing the network performance.
We evaluate via simulations the average throughput and outage probability of an indoor $ 60 $~GHz wireless communication system and quantify the impact of beam misalignment and beam pattern, and demonstrate the  trade-off in beam pattern design to balance the robustness against interference and beam misalignment.

%In contrast to the existing works, the specific contributions of our study can be summarized from the following aspects:
%\begin{enumerate}[label=(\emph{\roman*})]
%\item Different from the prevailing idealization of antenna pattern that consists of constant main-lobe and side-lobe gains, we consider a practical model where the main-lobe gain attenuates with the angle relative to the boresight. With the half-power beamwidth incorporated, the trade-off between the main-lobe and side-lobe gains are taken into account for analysis. 
%\item We introduce two beamwidth-dependent parameters, i.e., the half-power beamwidth ratio and misalignment deviation ratio, for characterizing the attenuation speed of main-lobe gain and the random beam misalignment, respectively. The impacts of these two parameters on the network performance are comprehensively investigated, and their great importance in network design is also quantified and highlighted.
%%\item We formulate the probability distribution of SINR when the random misalignment is considered, and the  analytical expression can be generally applied to a large variety of scenarios for performance analysis. 
%\item In addition to the formulation of probability distribution of SINR for general cases, we furthermore develop valid probabilistic bounds to keep track of the performance in the presence of finite concurrent transmissions in the area of interest, which largely facilitate the computation in characterizing the network performance.
%\end{enumerate}

The rest of the paper is organized as follows. We present the system model in Section~\ref{sec:models} and derive the probability distribution of SINR in the presence of random beam misalignment  in Section~\ref{sec:probabilistic analysis}. In Section~\ref{sec:example study}  we derive the bounds for the probability distribution of SINR performance. Performance evaluations are performed in Section~\ref{sec:performance evaluation} and conclusions are in Section~\ref{sec:conclusion}.

\section{System Model}\label{sec:models}

\subsection{Antenna Model with Beam Misalignment}
The 3GPP two-dimension directional antenna pattern~\cite{3gpp2014} is adopted in our study, where the antenna gain  $ G(\theta) $, with respect to the relative angle $ \theta $ to its boresight, is given by
\begin{equation}\label{eqn:3gpp antenna gain}
G\left(\theta\right)=
\begin{dcases}
G_m\cdot 10^{-\frac{3}{10}\left(\frac{2\theta}{\omega}\right)^2}, & |\theta| \leq \frac{\theta_{m}}{2},\\
G_{s}, & \frac{\theta_m}{2} \leq |\theta| \leq \pi,
\end{dcases}
\end{equation}
where  $ \omega $ denote the half-power (3 dB) beamwidth, and $ \theta_{m} $ is the main-lobe beamwidth.
$ G_{m} $ and $ G_{s} $ represent the maximum main-lobe gain and averaged side-lobe gain, respectively.
The total radiated power constraint~\cite{cai2010rex, wildman2014joint} requires that $\int_{-\pi}^{\pi}G(\theta)d\theta = 2\pi$, that is,
\begin{equation}\label{eqn:gains constraint 1}
\int_{0}^{\frac{\theta_m}{2}}G_m 10^{-\frac{3}{10}\left(\frac{2\theta}{\omega}\right)^2}d\theta + \int_{\frac{\theta_m}{2}}^{\pi}G_s d\theta =\pi,
\end{equation}
and the continuity of the radiation pattern \eqref{eqn:3gpp antenna gain} at the critical value $ \theta = \frac{\theta_m}{2}$ requires
\begin{equation}\label{eqn:gains constraint 2}
G_m = G_s \cdot 10^{\frac{3}{10}\left(\frac{\theta_m}{\omega}\right)^2}.
\end{equation}
Combining \eqref{eqn:gains constraint 1} and \eqref{eqn:gains constraint 2}, we can determine $ G_m $ and $ G_s $ analytically, in terms of $ \theta_m $ and $ \omega $, as
\begin{equation*}
\begin{dcases}
G_s =\frac{2\pi}{V\left(\theta_m,\omega\right)+2\pi - \theta_m}\\
G_m = \frac{2\pi \cdot 10^{\frac{3}{10}\left(\frac{\theta_m}{\omega}\right)^2}} {V\left(\theta_m,\omega\right)+2\pi - \theta_m}
\end{dcases},
\end{equation*}
where $ V\left(\theta_m,\omega\right) $ is given by
\begin{equation*}
V\left(\theta_m,\omega\right)=\int_{0}^{\theta_m} 10^{\frac{3}{10}\left(\frac{\theta_m^2 - \theta^2}{\omega^2}\right)}d\theta.
\end{equation*}

To highlight  the main-lobe radiation pattern,  we introduce  the parameter \emph{half-power to main-lobe beamwidth ratio}
\begin{equation}\label{eqn:eta}
\eta \triangleq \frac{\omega}{\theta_m} \in \left(0,1\right)
\end{equation}
to quantify the attenuation speed of the main beam gain. $\eta\to 1$ indicates an idealized constant-gain beam and $\eta\to 0$ mimics a fast-attenuating pencil beam.

Throughout the paper we assume that the random misalignment, denoted by ${\varepsilon} $,  is bounded within the range of the main-lobe beamwidth $ \theta_m $, namely, $ 0\leq |{\varepsilon}| \leq \frac{\theta_m}{2} $. This assumption is intuitively based on the fact that beam steering deviation exceeding the main-lobe beamwidth should be treated as \emph{alignment failure} rather than merely an \emph{misalignment}.  Furthermore, we assume that the misalignment $ \varepsilon$ follows a \emph{truncated normal distribution} with zero mean and variance $\sigma_{\varepsilon}^2$, that is,
\begin{equation}\label{eqn:pdf of misalignment}
f_{\varepsilon}(x)=\frac{\exp\left(-\frac{x^2}{2\sigma_{\varepsilon}^2}\right)}{\sigma_{\varepsilon} \sqrt{2\pi} \cdot \mathrm{erf}\left(\frac{\theta_m}{2\sqrt{2}\sigma_{\varepsilon}}\right)},\ |x|\leq \frac{\theta_m}{2},
\end{equation}
where $\mathrm{erf}\left(*\right) $ denotes the error function, and  $ \sigma_{\varepsilon} \in \left[0,\frac{\theta_m}{6}\right]$, i.e., $ 0\leq 3\sigma_{\varepsilon} \leq \frac{\theta_m}{2} $, mimicking the $ 3\sigma $-rule.
The \emph{misalignment deviation to main-lobe beamwidth ratio} is therefore defined as
\begin{equation}\label{eqn:MDMBR}
\rho \triangleq \frac{\sigma_{\varepsilon}}{\theta_{m}}\in \left[0,\frac{1}{6}\right].
\end{equation}

\subsection{Network Setting}\label{sec:network model}

\begin{figure}
\centering
\includegraphics[width=.65\columnwidth]{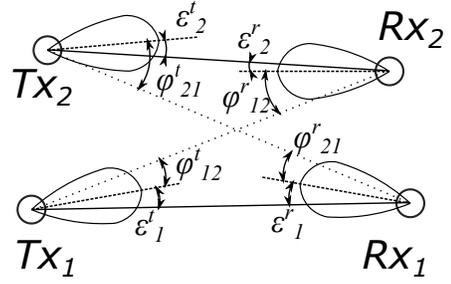}
\caption{Illustration of the two dimensional model with beam misalignment, concurrent transmission interference, and side-lobe signal leakage.}
\label{fig:network model with directional antennas}
\end{figure}

We consider a network that consists of $ N $ active communication pairs deployed randomly within an area of interest on a two dimensional plane, where for each communication pair $ i\in\{1,2,\ldots,N\} $,  the main beam of the transmitter $ \mathrm{TX}_{i} $ and the main beam of its intended receiver $ \mathrm{RX}_{i} $ are approximately aligned after appropriate channel/DoA estimation, position tracking, and beam steering. To highlight the impact of beam misalignment and to simplify presentation, we assume that all the transmitters and receivers have the same antenna radiation pattern as described in \eqref{eqn:3gpp antenna gain}, and extension to heterogeneous antenna patterns is straightforward.
In Fig.~\ref{fig:network model with directional antennas} we illustrate a snapshot of the beam misalignment and concurrent transmission interference between two neighboring communication pairs.
We denote by $ \varepsilon_{i}^{t} $ and $ \varepsilon_{i}^{r} $ the beam alignment errors (i.e., the angle between the transmission path\footnote{Here we assume line-of-sight (LOS) transmission in a short distance where the ``optical'' LOS path provides the highest gain (i.e., lowest loss). Otherwise the solid lines represent the logical LOS paths that provide the highest gain.} and misaligned boresight) of the $ i^{\mathrm{th}} $ link at the transmitter and the receiver sides, respectively. %\footnote{We assume the random misalignment is negative or positive when the boresight biases to right-hand side or left-hand side of the LOS path, correspondingly.}.
The incident angle of interference (with respect to the boresight of the receiver) from $ \mathrm{TX}_{j} $ to $ \mathrm{RX}_{i} $, $ i\neq j $, is denoted by $ \varphi_{ji}^{r} $, and the departure angle of interference (w.r.t. the boresight of the transmitter) is represented by $ \varphi_{ji}^{t} $.

The desired signal strength can therefore be represented as a function of the  beam alignment errors $ \varepsilon_{i}^{t} $ and $ \varepsilon_{i}^{r} $, and the interference power can be written as a function of the incident angles $ \varphi_{ji}^{r} $ and  $ \varphi_{ji}^{t}$.
 The SINR at receiver $ \mathrm{RX}_{i} $ is written as
\begin{equation}\label{eqn:SINR}
\gamma_{i}\triangleq \frac{P_{r,i}}{N_{0}+I_{i}}= \frac{P_{t}G\left(\varepsilon_{i}^{t}\right)G\left(\varepsilon_{i}^{r}\right)L\left(d_{i}\right)}{N_{0}+P_{t}\sum\limits_{k\neq i}G\left(\varphi_{ki}^{t}\right)G\left(\varphi_{ki}^{r}\right)L\left(d_{ki}\right)},
\end{equation}
where $P_t$ is the transmit power,  $ N_{0} $ is the noise power, and $G(\theta)$ represents the antenna gain with respect to angle $ \theta $. $ P_{r,i} $ represents the power of the received signal, $ I_{i} $ is the aggregate interference power at $ \mathrm{RX}_{i} $,  $d_{i}$ is the transmission distance from $\mathrm{TX}_{i} $ to $ \mathrm{RX}_{i}$, and $d_{ki}$ is the distance from $\mathrm{TX}_{k} $ to $\mathrm{RX}_{i}$, $ k\neq i $. $L(d)$ denotes the path loss at distance $d$, which is given by
$$ L\left(d\right) = \left(\frac{\lambda}{4\pi}\right)^2 d^{-\alpha}, $$
where $\lambda$ is the carrier wavelength, and $\alpha$ is the path loss attenuation exponent. We assume that $ d\geq d_0 = 0.5 $ meter to ensure the far field for radio propagation.

\section{Beam Misalignment and Interference}\label{sec:probabilistic analysis}

When the mobility of user terminals is small, the SINR observed during a small period of time relies on the positions of all the active nodes. We describe the positions of an active communication pair in the two dimensional plane  by a complex vector  $ \mathbf{Q}_{i}=\left[Q_{i}^{t},Q_{i}^{r}\right]^{T}\in \mathbb{C}^2 $, where $ Q_{i}^{t} $ and $ Q_{i}^{r} $ represent the location information of $ \mathrm{TX}_{i} $ and $ \mathrm{RX}_{i} $, respectively.
Likewise, all the neighboring concurrent transmissions can be captured by vectors $ \mathbf{Q}_{j} $, $ j\neq i $, based on which the aggregate interference $I_{i}$ can be computed. For the sake of simplicity, we take the node pair $ \left(\mathrm{TX}_1,\mathrm{RX}_1\right) $ as the typical object for investigation.

It is worth pointing out that, the received signal power $ P_{r,1} $ depends on both $ \varepsilon_1^r $ and  $ \varepsilon_1^t $, and the interference power $ I_{1} $ depends on $ \varepsilon^r_1 $ and  $ \varepsilon^t_j $, for $ j=2,...,n $. Therefore, $ P_{r,1} $ is correlated with $ I_{1} $ through $ \varepsilon_1^r $. Given the set of $ n $ random location information vectors, namely, $ \mathbf{Q}^{(n)}\triangleq \left(\mathbf{Q}_{1},\mathbf{Q}_{2},\ldots,\mathbf{Q}_{n}\right) $ and the beam misalignment $ \varepsilon_{1}^{r} $ at $ \mathrm{RX}_{1} $, the probability density function (p.d.f.) of the SINR $ \gamma_{1} $ can be expressed as
\begin{equation}\label{eqn:pdf of SINR}
\begin{split}
 f_{\gamma_{1}}\left(x\right)= \idotsint  &  f_{\gamma_{1}|\mathbf{Q}^{(n)},\varepsilon_{1}^{r}}\left(x|\mathbf{q}^{(n)},e\right)\\ & f_{\mathbf{Q}^{(n)},\varepsilon_{1}^{r}}\left(\mathbf{q}^{(n)},e\right) d\mathbf{q}_1 \dots d\mathbf{q}_n de,
\end{split}
\end{equation}
where $ f_{\gamma_{1}|\mathbf{Q}^{(n)},\varepsilon_{1}^{r}}\left(x|\mathbf{q}^{(n)},e\right) $ is the conditional p.d.f. of $ \gamma_{1} $ given $\left( \mathbf{Q}^{(n)},\varepsilon_{1}^{r}\right)=\left(\mathbf{q}^{(n)},e\right) $, with $ \mathbf{q}^{(n)}\triangleq \left(\mathbf{q}_{1},\mathbf{q}_{2},\ldots,\mathbf{q}_{n}\right) $. $ f_{\mathbf{Q}^{(n)},\varepsilon_{1}^{r}}\left(\mathbf{q}^{(n)}, e\right) $ denotes the joint p.d.f. of $ \left(\mathbf{Q}^{(n)},\varepsilon_{1}^{r}\right)$, which can be reduced to (due to the independence of $ \mathbf{Q}^{(n)} $ and $ \varepsilon_{1}^{r} $)
\begin{equation*}
f_{\mathbf{Q}^{(n),\varepsilon_{1}^{r}}}\left(\mathbf{q}^{(n)},e\right) = f_{\mathbf{Q}^{(n)}}\left(\mathbf{q}^{(n)}\right)f_{\varepsilon_{1}^{r}}\left(e\right).
\end{equation*}

%The conditional p.d.f. $ f_{\gamma_{1}|\mathbf{Q}^{(n)},\varepsilon_{1}^{r}}\left(x|\mathbf{q}^{(n)},e\right) $ can be determined according to \eqref{eqn:SINR} by deriving the conditional probability functions of signal power $ P_{r,1} $ and interference power $ I_{1} $.

\begin{proposition}\label{proposition:pdf of SINR}
Let $ \mathbf{Q}^{(n)}$ and $ \varepsilon_{1}^{r} $ be the set of random location information vectors for $ n $ links and the beam misalignment at $ \mathrm{RX}_{1} $, respectively, the conditional p.d.f. of SINR $ \gamma_{1} $ by \eqref{eqn:SINR} given $ \mathbf{Q}^{(n)}=\mathbf{q}^{(n)} $ and $ \varepsilon_{1}^{r}=e $ is obtained as 
\begin{align}
& f_{\gamma_{1}|\mathbf{Q}^{(n)},\varepsilon_{1}^{r}}\left(x|\mathbf{q}^{(n)},e\right)  = \label{eqn:proposition1} \\
&\int_{N_{0}}^{\infty}y f_{P_{r,1}|\mathbf{Q}_{1},\varepsilon_{1}^{r}}\left(xy|\mathbf{q}_{1},e\right)  
  f_{I_{1}|\mathbf{Q}^{(n)},\varepsilon_{1}^{r}}\left(y-N_{0}|\mathbf{q}^{(n)},e\right)dy, \nonumber
\end{align} 
where $ f_{P_{r,1}|\mathbf{Q}_{1},\varepsilon_{1}^{r}}\left(*|\mathbf{q}_{1},e\right) $ and $ f_{I_{1}|\mathbf{Q}^{(n)},\varepsilon_{1}^{r}}\left(*|\mathbf{q}^{(n)},e\right) $ denote the conditional p.d.f. of $ P_{r,1} $ and $ I_{1} $, respectively.
\end{proposition}
\begin{IEEEproof}
Given two independent positive random variables $ Y $ and $ W $  with p.d.f. $ f_Y(y) $ and $ f_{W}(w) $, respectively,  by applying the p.d.f. computation for the product of two random variables (see Appendix), it is straightforward to derive the p.d.f. of
\begin{equation*}
X \triangleq \frac{Y}{c+W} = Y\cdot\left(c+W\right)^{-1}, 
\end{equation*}
where $ c $ is a positive constant. Note that $ P_{r,1} $ and $ I_{1} $ are conditionally independent given $ \mathbf{Q}^{(n)}=\mathbf{q}^{(n)} $ and $ \varepsilon_{1}^{r}=e $, we have
\begin{equation*}
\begin{split}
 f_{\gamma_{1}|\mathbf{Q}^{(n)},\varepsilon_{1}^{r}}\left(x|\mathbf{q}^{(n)},e\right)
= & \int_{N_{0}}^{\infty}y f_{P_{r,1}|\mathbf{Q}^{(n)},\varepsilon_{1}^{r}}\left(xy|\mathbf{q}^{(n)},e\right) \\ & \cdot f_{I_{1}|\mathbf{Q}^{(n)},\varepsilon_{1}^{r}}\left(y-N_{0}|\mathbf{q}^{(n)},e\right)dy.
\end{split}
\end{equation*}

Since $ P_{r,1} $ depends on $ \mathbf{Q}^{(n)}=\mathbf{q}^{(n)} $ only through $ \mathbf{Q}_{1}=\mathbf{q}_{1}$, the p.d.f. of SINR  $\gamma_1$  can be obtained as \eqref{eqn:proposition1}.
%Back to our specific problem, it is easy to prove that, $ P_{r,1} $ is conditional independent with $ I_{1} $, given the condition $ \mathbf{Q}^{(n)}=\mathbf{q}^{(n)} $, and $ \varepsilon_{1}^{r}=e $, and $ \mathbf{Q}^{(n)}=\mathbf{q}^{(n)} $ can be further reduced to $ \mathbf{Q}_{1}=\mathbf{q}_{1} $ for $ P_{r,1} $. That equivalently indicates $ Y $ and $ W $ here can be treated as independent random variables.
\end{IEEEproof}

%Through Proposition \ref{proposition:pdf of SINR}, we know that conditional p.d.f.s of $ P_{r,1} $ and $ I_{1} $ given $ \mathbf{Q}^{(n)}=\mathbf{q}^{(n)} $  and $ \varepsilon_{1}^{r}=e $ are crucial for finally obtaining the p.d.f. of SINR $ \gamma_{1} $. Additionally, due to the fact of conditional independence between $ P_{r,1} $ and $ I_{1} $, we can then separately study their conditional p.d.f.s.

\subsection{Distribution of Signal Power with Beam Misalignment}\label{sec:Pr}
%As mentioned in Proposition~\ref{proposition:pdf of SINR}, for the received power $ P_{r,1} $, the conditional hypothesis $ \mathbf{Q}^{(n)}=\mathbf{q}^{(n)} $ can be reduced to $ \mathbf{Q}_{1}=\mathbf{q}_{1} $. More precisely, l
Let $ P_{t} $ denote the transmit signal power and assume that the transmit beam gain $ g_{\varepsilon_{1}^{t}} =  G\left(\varepsilon_{1}^{t}\right) $ is a random variable with associated p.d.f. $ f_{g}\left(x\right) $, $ x\in \left[G_s,G_m\right] $. The received signal power $ P_{r,1} $ given $ \mathbf{Q}_{1}=\mathbf{q}_{1}\triangleq \left[q_{1}^{t},q_{1}^{r}\right]^{T} \in \mathbb{C}^{2}$ and $ \varepsilon_{1}^{r} =e$ can be reformulated as
\begin{equation}\label{eqn:conditional P_r1}
\begin{split}
P_{r,1}\big|_{\mathbf{Q}_{1} = \mathbf{q}_{1},\varepsilon_{1}^{r} = e}
= & P_{t} L\left(d_{11}\right)G\left(e\right)\cdot g_{\varepsilon_{1}^{t}},
\end{split}
\end{equation}
where $ d_{11}{\triangleq} |q_{1}^{r}{-}q_{1}^{t}| $ represents the length of the link.
The conditional p.d.f. $ f_{P_{r,1}|\mathbf{Q}_{1},\varepsilon_1^{r}}\left(x|\mathbf{q}_{1},e\right) $ can therefore be determined  by the p.d.f. of $ g_{\varepsilon_{1}^{t}}$ as shown below.% in Proposition~\ref{proposition:pdf of received power}.

\begin{proposition}\label{proposition:pdf of received power}
Let $ f_{\varepsilon_1^t}\left(y\right)$, $|y|{\leq} \theta_m/2$ be the p.d.f. of beam misalignment $ \varepsilon_{1}^{t}$, the conditional p.d.f. $ f_{P_{r,1}|\mathbf{Q}_{1},\varepsilon_1^r}\left(x|\mathbf{q}_{1},e\right) $ given $ \mathbf{Q}_{1}{=}\mathbf{q}_{1} $ and $ \varepsilon_1^r{=}e $ is written as
\begin{equation*}
\begin{split}
& f_{P_{r,1}|\mathbf{Q}_{1},\varepsilon_1^r}\left(x|\mathbf{q}_{1},e\right)=\frac{1}{P_{t} L\left(d_{11}\right)g_e} f_{g_{\varepsilon_1^t}}\left(\frac{x}{P_{t} L\left(d_{11}\right)g_e}\right),
\end{split}
\end{equation*}
where $ g_e {=} G\left(e\right) $ as described by the radiation pattern \eqref{eqn:3gpp antenna gain} and the p.d.f. $f_{g_{\varepsilon_1^t}}\left(x\right)$ for $ \ x\in \left[G_s, G_{m}\right] $ can be written as
\begin{equation}\label{prop:2-1}
f_{g_{\varepsilon_1^t}}(x)=\frac{\omega f_{\varepsilon_1^t}\left(\omega\sqrt{\frac{5}{6}\log_{10}\left(\frac{G_{m}}{x}\right)}\right)}{\ln(10) x\sqrt{\frac{6}{5}\log_{10}\left(\frac{G_{m}}{x}\right)}}.
\end{equation}
\end{proposition}
\begin{IEEEproof}
Since $ 0\leq e\leq \theta_m/2$, for  $ g_{e}= G\left(e\right) $ we can derive from   \eqref{eqn:3gpp antenna gain} that
\begin{equation*}
e = \frac{\omega}{2}\sqrt{\frac{10}{3}\log_{10}\left(\frac{G_{m}}{g_{e}}\right)}.
\end{equation*}
Note that the function $ g_{e}=G(e) $ is differentiable within the interval $ 0\leq e < \theta_m/2$, we have
\begin{equation*}
G'(e)=-\frac{12\ln\left(10\right)}{5\omega^2} e g_{e} =-\frac{2\ln\left(10\right)}{\omega}g_{e}\sqrt{\frac{6}{5}\log_{10}\left(\frac{G_{m}}{g_{e}}\right)},
\end{equation*}
and the p.d.f. of $ g_{e} $ can be straightforwardly derived from the p.d.f. $f_{\epsilon}(e)$ given in \eqref{eqn:pdf of misalignment}, as shown in \eqref{prop:2-1}. We can now apply \eqref{prop:2-1} to  \eqref{eqn:conditional P_r1} to conclude the proof.
%As discussed in \eqref{eqn:conditional P_r1}, the problem can be reformulated as a function of the p.d.f. of $ \varepsilon_{1}^{t} $, with the constant scalar $ P_{t}L\left(d_{11}\right)g_e $ given by conditions $ \mathbf{Q}_{1}=\mathbf{q}_{1} $ and $ \varepsilon_1^r=e $. Regarding the p.d.f. $ f_{g_{\varepsilon_1^t}}\left(*\right) $, we define the function $ g_{\varepsilon_1^t}=G\left(\varepsilon_1^t\right) $ by following \eqref{eqn:3gpp antenna gain}, hence the p.d.f. of $ g_{\varepsilon_1^t} $ can be easily obtained with respect to the p.d.f. of $ \varepsilon_1^t $.
\end{IEEEproof}
%
%The p.d.f. of random beam misalignment $ \varepsilon $, $ |\varepsilon|\leq \bar{\varepsilon} $, is written as $ f_{\varepsilon}(x) $.

\subsection{Distribution of Interference Power}
Let $ I_{1} = \sum_{j=2}^{n}I_{j1}$ be the sum interference power where  $ I_{j1} $ is the interference power from the $ j^{\mathrm{th}} $ concurrent transmission to $ \mathrm{RX}_{1} $. In Lemma~\ref{lemma:conditional independence of interference}, we show that $ I_{j1} $, $ j=2,3,\ldots,n $ are conditional independent given $ \mathbf{Q}^{(n)} = \mathbf{q}^{(n)}$ and $ \varepsilon_1^r=e $.

\begin{lemma}\label{lemma:conditional independence of interference}
Let $ I_{j1} $, $ j=2,3,\ldots,n $, denote the interference power to $ \mathrm{RX}_{1} $ from $ \mathrm{TX}_{j} $, the conditional joint p.d.f. $ f_{I_{21},\ldots,I_{n1}|\mathbf{Q}^{(n)},\varepsilon_1^r}\left(x_{2},\ldots,x_{n}|\mathbf{q}^{(n)}, e\right) $ can be written as
\begin{equation*}
\begin{split}
& f_{I_{21},\ldots,I_{n1}|\mathbf{Q}^{(n)},\varepsilon_1^r}\left(x_{2},\ldots,x_{n}|\mathbf{q}^{(n)},e\right) \\ = & \prod\limits_{j=2}^{n}f_{I_{j1}|\mathbf{Q}_{1},\mathbf{Q}_{j}, \varepsilon_1^r}\left(x_{j}|\mathbf{q}_{1},\mathbf{q}_{j}, e\right),
\end{split}
\end{equation*}
where $ f_{I_{j1}|\mathbf{Q}_{1},\mathbf{Q}_{j}, \varepsilon_1^r}\left(*|\mathbf{q}_{1},\mathbf{q}_{j}, e\right) $ , $ j=2,\ldots,n $, is the conditional p.d.f. of $ I_{j1} $ given both $ \mathbf{Q}_{1}=\mathbf{q}_{1} $, $ \mathbf{Q}_{j}=\mathbf{q}_{j} $ and $ \varepsilon_1^r = e $.
\end{lemma}
\begin{IEEEproof} 
Given $ \mathbf{Q}^{(n)}=\mathbf{q}^{(n)} $ and $ \varepsilon_1^r = e $, it is easy to obtain that 
\begin{equation*}
\begin{split}
& f_{I_{21},\ldots,I_{n1}|\mathbf{Q}^{(n)},\varepsilon_1^r}\left(x_{2},\ldots,x_{n}|\mathbf{q}^{(n)}, e\right) \\
\overset{(a)}{=} & \prod\limits_{j=2}^{n}f_{I_{j1}|I_{(j+1)1},\ldots,I_{n1},\mathbf{Q}^{(n)},\varepsilon_1^r}\left(x_{j}|x_{(j+1)},\ldots,x_{n},\mathbf{q}^{(n)}, e\right) \\
\overset{(b)}{=} & \prod\limits_{j=2}^{n}f_{I_{j1}|\mathbf{Q}^{(n)}, \varepsilon_1^r }\left(x_{j}|\mathbf{q}^{(n)}, e\right)\\
\overset{(c)}{=} & \prod\limits_{j=2}^{n}f_{I_{j1}|\mathbf{Q}_{1},\mathbf{Q}_{j}, \varepsilon_1^r }\left(x_{j}|\mathbf{q}_{1},\mathbf{q}_{j}, e\right)
,
\end{split}
\end{equation*}
where $ (a) $ applies the chain rule of conditional p.d.f. for multivariate random variables, $ (b) $ comes from the fact that $ I_{j1}--(\mathbf{Q}^{(n)}, \varepsilon_1^r)--I_{j' 1} $ forms a markov chain  for all $ j'\neq j $,  $ (c) $  is due to the dependence of $ I_{j1}$ on   $ \mathbf{Q}^{(n)}=\mathbf{q}^{(n)} $ only through the the pair $ \left(\mathrm{TX}_{j}, \mathrm{RX}_{1}\right)$, which  
reduces the condition $ \mathbf{Q}^{(n)}=\mathbf{q}^{(n)} $  to $ \mathbf{Q}_{1}=\mathbf{q}_{1} $ and $ \mathbf{Q}_{j}=\mathbf{q}_{j} $.
\end{IEEEproof}

%The conditional independence by Lemma~\ref{lemma:conditional independence of interference} allows analyzing $ f_{I_{j1}|\mathbf{Q}_{1},\mathbf{Q}_{j}, \varepsilon_1^r }\left(*|\mathbf{q}_{1},\mathbf{q}_{j}, e\right) $ independently. 

Since the component interference $ I_{j1} $ given $ \mathbf{Q}^{(n)} {=} \mathbf{q}^{(n)}$ and $\varepsilon_{1}^{r} {=} e $ can be reformulated as
\begin{equation}\label{eqn:conditional I_j1}
I_{j1}\big|_{\mathbf{Q}^{(n)} = \mathbf{q}^{(n)},\varepsilon_{1}^{r} = e}
=  P_{t} L\left(d_{j1}\right)G\left(\varphi_{j1}^{r}\right)\cdot g_{\varphi_{j1}^{t}},
\end{equation}
where $ d_{j1}\triangleq |q_{1}^{r}-q_{j}^{t}| $ is the distance between $ \mathrm{TX}_{j} $ and $ \mathrm{RX}_{1} $, and $ g_{\varphi_{j1}^{t}} \triangleq G\left(\varphi_{j1}^{t}\right) $ is a function of random variable $ \varphi_{j1}^{t} $,  we will establish in Lemma~\ref{lemma:pdf of AoD} the conditional p.d.f. of $ \varphi_{j1}^{t} $.

\begin{lemma}\label{lemma:pdf of AoD}
%Regarding the interfering link $ \left(\mathrm{TX}_{j},\mathrm{RX}_{1}\right) $,
Given $ \mathbf{Q}^{(n)}=\mathbf{q}^{(n)} $, the departure angle $ \varphi_{j1}^{t}\in \left[0,\pi\right]$ of the interfering link $ \left(\mathrm{TX}_{j},\mathrm{RX}_{1}\right) $ can be written as
\begin{equation}\label{lemma_21}
\begin{split}
\varphi_{j1}^{t}\big|_{\mathbf{Q}^{(n)}= \mathbf{q}^{(n)}} \!
= \!
\begin{dcases}
|2\pi - |\hat{\varphi}_{j1}^{t}-\varepsilon_j^t||, & |\hat{\varphi}_{j1}^{t}-\varepsilon_j^t| \geq \pi,\\
|\hat{\varphi}_{j1}^{t}- \varepsilon_j^t|, & \mbox{otherwise},
\end{dcases}
\end{split}
\end{equation}
where $ \hat{\varphi}_{j1}^{t}\triangleq \angle\left(\frac{q_j^r-q_j^t}{q_1^r-q_j^t}\right)\in \left[-\pi,\pi\right) $ represents the signed angle\footnote{Given two complex variables $ u_1 $ and $ u_2 $, the signed angle $ \angle\left(\frac{u_1}{u_2}\right) $ denotes the rotated angle from $ u_1 $ to $ u_2 $, which is defined to be negative if the rotation occurs in the clockwise direction.} under perfect beam alignment given $ \mathbf{Q}^{(n)}=\mathbf{q}^{(n)} $.
Its conditional p.d.f. $ f_{\varphi_{j1}^t|\mathbf{Q}^{(n)}}\left(*|\mathbf{q}^{(n)}\right) $ is given by
\begin{align}
& f_{\varphi_{j1}^t|\mathbf{Q}^{(n)}}\left(x|\mathbf{q}^{(n)}\right)
=  \left(1-F_{z|\mathbf{Q}^{(n)}}\left(2\pi\right)\right)f_{z|\mathbf{Q}^{(n)}}\left(x+2\pi\right) \nonumber\\
& + \left(F_{z|\mathbf{Q}^{(n)}}\left(2\pi\right)-F_{z|\mathbf{Q}^{(n)}}\left(\pi\right)\right)f_{z|\mathbf{Q}^{(n)}}\left(2\pi-x\right) \label{eqn:lemma-12}\\
& + F_{z|\mathbf{Q}^{(n)}}\left(\pi\right)f_{z|\mathbf{Q}^{(n)}}\left(x\right), \nonumber
\end{align}
where $ f_{z|\mathbf{Q}^{(n)}}\left(*\right) $ and $ F_{z|\mathbf{Q}_{(n)}}\left(*\right) $  denote the conditional p.d.f. and c.d.f., respectively,  of $ z = |\hat{\varphi}_{j1}^t-\varepsilon_j^t|$, with %and $ f_{z|\mathbf{Q}^{(n)}}\left(*\right) $  is given by
\begin{equation}
f_{z|\mathbf{Q}^{(n)}}\left(x\right) = f_{\varepsilon_j^t}\left(\hat{\varphi}_{j1}^t +x\right) + f_{\varepsilon_j^t}\left(\hat{\varphi}_{j1}^t -x\right).
\end{equation}
\end{lemma}
%Let $ f_{\varepsilon_j^t}\left(*\right) $ be the p.d.f. of random beam misalignment $ \varepsilon_j^t\geq 0 $,
\begin{IEEEproof}
In the absence of beam misalignment, the angle $ \hat{\varphi}_{j1}^{t} $ that represents the angle-of-departure (AoD) at $ \mathrm{TX}_{j} $ is determined by $ \mathbf{Q}^{(n)}{=}\mathbf{q}^{(n)} $. The AoD with misalignment, denoted as $ \varphi_{j1}^{t} $, is the sum of the deterministic $ \hat{\varphi}_{j1}^{t} $ and a stochastic $ \varepsilon_{j}^{t} $, as modeled in \eqref{lemma_21}.
%
%Thus, we immediately come to \eqref{eqn:lemma-11}.
Setting $ z {=} | \hat{\varphi}_{j1}^t - \varepsilon_j^t |$, its conditional c.d.f.  $ F_{z|\mathbf{Q}^{(n)}}\left(t\right) $ can be expressed as
\begin{equation*}
F_{z|\mathbf{Q^{(n)}}}\left(t|\mathbf{q}^{(n)}\right)  = \mathbb{P}\left(| \hat{\varphi}_{j1}^t - \varepsilon_j^t |\leq t\right),
\end{equation*}
from which the conditional p.d.f. $ f_{z|\mathbf{Q}_{(n)}}\left(t|\mathbf{q}_{(n)}\right) $ can be obtained. Furthermore, we have
\begin{equation*}
\begin{split}
& F_{\varphi_{j1}^t|\mathbf{Q}^{(n)}}\left(y|\mathbf{q}^{(n)}\right)
= \mathbb{P}\left(z - 2\pi \leq y \right)\mathbb{P}\left(z\geq 2\pi \right) \\
& + \mathbb{P}\left(2\pi-z \leq y \right)\mathbb{P}\left(\pi \leq z < 2\pi \right)
  + \mathbb{P}\left(z\leq y\right)\mathbb{P}\left(z< \pi\right).
\end{split}
\end{equation*}
Taking the first derivative of $ F_{\varphi_{j1}^t|\mathbf{Q}^{(n)}}\left(y|\mathbf{q}^{(n)}\right) $ with respect to $ y $ leads to \eqref{eqn:lemma-12}.% and concludes the lemma by deriving the conditional p.d.f. of $ \varphi_{j1}^t $.
\end{IEEEproof}

%\begin{lemma}\label{lemma:pdf of AoA}
Likewise, the arrival angle $ \varphi_{j1}^{r}\in \left[0,\pi\right] $  of the interfering link $ \left(\mathrm{TX}_{j},\mathrm{RX}_{1}\right)$  given $ \mathbf{Q}^{(n)}{=}\mathbf{q}^{(n)} $ and $ \varepsilon_1^r {=} e $ is written as
\begin{equation}\label{lemma:pdf of AoA}
\varphi_{j1}^{r}\big|_{\mathbf{q}^{(n)}, e} =
\begin{dcases}
|2\pi - |\hat{\varphi}_{j1}^{r}-e||, & |\hat{\varphi}_{j1}^{r} - e| \geq \pi,\\
|\hat{\varphi}_{j1}^{r} - e|, & \mbox{otherwise},
\end{dcases}
\end{equation}
where $ \hat{\varphi}_{j1}^{r}\triangleq \angle\left(\frac{q_1^t-q_1^r}{q_j^t-q_1^r}\right)\in \left[-\pi,\pi\right) $ is the angle corresponding to the perfect beam alignment given $ \mathbf{Q}^{(n)}{=}\mathbf{q}^{(n)} $.
%\end{lemma}
%Its conditional p.d.f. The statement of Lemma~\ref{lemma:pdf of AoA} is similar to that in Lemma~\ref{lemma:pdf of AoD}.

%In the light of Lemma~\ref{lemma:pdf of AoD} and Lemma~\ref{lemma:pdf of AoA}, similar to \eqref{eqn:conditional P_r1},

\begin{proposition}\label{proposition:pdf of component interference}
The p.d.f. of $I_{j1}$ given $\mathbf{Q}_1,\mathbf{Q}_j,\varepsilon_1^r$ is
%The conditional p.d.f. $ f_{I_{j1}|\mathbf{Q}_1,\mathbf{Q}_j,\varepsilon_1^r}\left(x|\mathbf{q}_1,\mathbf{q}_j, e\right) $  is
\begin{align}
& f_{I_{j1}|\mathbf{Q}_1,\mathbf{Q}_j,\varepsilon_1^r}\left(x|\mathbf{q}_1,\mathbf{q}_j, e\right) \label{eqn:condition reduction for interference}\\
= & \frac{1}{P_{t} L\left(d_{j1}\right)g_{\varphi_{j1}^{r}}} f_{g_{\varphi_{j1}^{t}}|\mathbf{Q}_1,\mathbf{Q}_j}
\bigg(\frac{x}{P_{t} L\left(d_{j1}\right)g_{\varphi_{j1}^{r}}}\big|\mathbf{q}_1,\mathbf{q}_j \bigg), \nonumber
\end{align} 
where,  for $ x\in \left[G_{s}, G_{m}\right] $, we have
\begin{equation*}
\begin{split}
f_{g_{\varphi_{j1}^{t}}|\mathbf{Q}_1,\mathbf{Q}_j }\left(x|\mathbf{q}^{(n)}\right)
=  \frac{\omega f_{\varphi_{j1}^{t}|\mathbf{Q}^{(n)}}\left(\omega\sqrt{\frac{5}{6}\log_{10}\left(\frac{G_{m}}{x}\right)}\Big|\mathbf{q}^{(n)}\right)}{\ln(10) x\sqrt{\frac{6}{5}\log_{10}\left(\frac{G_{m}}{x}\right)}}.
\end{split}
\end{equation*} 
\end{proposition}
\begin{IEEEproof}
By \eqref{eqn:conditional I_j1}, \eqref{lemma:pdf of AoA}, and Lemma~\ref{lemma:pdf of AoD}, it is straightforward to obtain 
the results by applying the similar method as shown in Proposition~\ref{proposition:pdf of received power}. 
\end{IEEEproof}

We can now derive the conditional p.d.f. of $ I_{1} $ as follows.%, presented in Proposition~\ref{proposition:aggregated interference}.

\begin{proposition}\label{proposition:aggregated interference}
%Let $ \mathbf{Q}^{(n)} $ be the set of random location information vectors for $ n $ links, and denote by $ I_{j1} $, $ j=2,3,\ldots,n $, the interfering power to the typical receiver $ \mathrm{RX}_{1} $ from the interferer $ \mathrm{TX}_{j} $, then
The conditional p.d.f. of the sum interference given $ \mathbf{Q}^{(n)}= \mathbf{q}^{(n)}$ and $ \varepsilon_1^r=e $ is given by
\begin{equation*}
f_{I_{1}|\mathbf{Q}^{(n)},\varepsilon_1^r}\left(x|\mathbf{q}^{(n)},e\right) = \bigotimes_{j=2}^{n}f_{I_{j1}|\mathbf{Q}_1,\mathbf{Q}_j,\varepsilon_1^r}\left(x|\mathbf{q}_1,\mathbf{q}_j,e\right),
\end{equation*}
where %$ f_{I_{j1}|\mathbf{Q}_1,\mathbf{Q}_j,\varepsilon_1^r}\left(x|\mathbf{q}_1,\mathbf{q}_j,e\right) $ refers to Proposition~\ref{proposition:pdf of component interference}, and 
$ \bigotimes $ represents the convolution operator.
\end{proposition}
\begin{IEEEproof}
Since $ I_{j1} $, $ j{=}2,3,\ldots,n $, are conditionally independent given $ \mathbf{Q}^{(n)}{=}\mathbf{q}^{(n)} $ and $ \varepsilon_1^r{=}e $, the p.d.f. of the sum of independent random variables equals the convolution of all the individual probability functions.
\end{IEEEproof}

Finally, the conditional p.d.f. of SINR in Proposition~\ref{proposition:pdf of SINR} can be obtained by applying  Proposition~\ref{proposition:pdf of received power} and Proposition~\ref{proposition:aggregated interference}, which is then used to compute the p.d.f. of SINR using \eqref{eqn:pdf of SINR}.

%It is evident that, the above expressions are derived in a generalized sense, such that it is valid for any kind of antenna patterns, bounded beam misalignments, and node distributions. To highlight our work, we in the subsequent section take a specific scenario as an example to analyze the performance. 

\section{CDF of SINR: Upper Bound and Lower Bound}\label{sec:example study}
It is rather involved to directly evaluate the SINR performance based on the equations derived in the Sec.~\ref{sec:probabilistic analysis}, partially due to the convolution of p.d.f. in Proposition~\ref{proposition:aggregated interference}. 
For scenarios where there are $K$ interfering transmitters distributed uniformly at random around the receiving node $\mathrm{RX}_1$, whose location  $\mathbf{Q}_1 = \mathbf{q}_1$  and beam misalignment $\varepsilon_1^r=e$ are given, we derive upper and lower bounds on the c.d.f. of SINR for  $\mathrm{RX}_1$.
According to Lemma~\ref{lemma:conditional independence of interference}, we know that,  in the presence of the given $\mathbf{Q}_1$ and $\varepsilon_1^r$, the component interference power $I_{j1}$, $j\in \{2,\ldots,K+1\}$, can be treated as independent random variables. Furthermore,  $I_{j1}$ are also identically distributed random variables due to the uniform deployments and orientations. Thus, the interference $ I_{j1} $ can be viewed as independent and identically distributed (i.i.d.) random variables.

Following Lemma~\ref{lemma:conditional independence of interference} and Proposition~\ref{proposition:pdf of component interference}, we can obtain the conditional probability $ f_{I_{j1}|\mathbf{Q}_1,\varepsilon_1^r } \left(*|\mathbf{q}_1,e\right)$ by marginalizing out the variable $ \mathbf{Q}_j $, which covers the location information of the $ j^{\mathrm{th}} $ transmission pair. Since only $ Q_j^t $ is required for the marginalization process, we have
\begin{align*}
f_{I_{j1}|\mathbf{Q}_1,\varepsilon_1^r } \left(x|\mathbf{q}_1, \! e\right) \!
= \! & \int \!\! f_{I_{j1}|\mathbf{Q}_1, Q_j^t,\varepsilon_1^r }\! \left(x|\mathbf{q}_1,q_j^t,e\right)f_{Q_j^t} \! \left(q_j^t\right) dq_j^t. 
\end{align*}

  For notational simplicity, we use $ Y $ and $ W_j $, $ 2\leq j \leq K+1 $, respectively to represent the random variables $ P_{r,1} $ and $ I_{j1} $ conditional on $ \mathbf{Q}_1= \mathbf{q}_1$ and $ \varepsilon_1^r=e $. We can then rewrite the conditional c.d.f. $ \mathbb{P}\left(\gamma_1\leq x | \mathbf{q}_1,e\right) $ as
\begin{equation}
\mathbb{P}\left(\gamma_1\leq x | \mathbf{q}_1,e\right)  \triangleq \mathbb{P}\left(\frac{Y}{N_0 + W_{\sum}}\leq x\right),
\end{equation}
where  $W_{\sum} \triangleq \sum_{j=2}^{K+1} W_j$.
%
%\subsection{Bound Analysis}
%
Denote the c.d.f. of $Y$ and $W_{\sum}$  by $F_{Y}\left(*\right)$ and $ F_{W_{\sum}}\left(*\right) $, respectively.  $F_{Y}\left(x\right)$ can be immediately obtained by applying Proposition~\ref{proposition:pdf of received power}, i.e., $F_{Y}\left(x\right) = \int_0^{x}f_{Y}\left(t\right)dt$. For the sum interference $W_{\sum}$ with respect to $K$ ($K\geq 1$) interfering transmitters, from Proposition~\ref{proposition:aggregated interference}, we know that
\begin{align*}
F_{W_{\sum}}(x)= \int_0^x f_{W_{\sum}}(t)dt = \int_0^x \bigotimes_{j=2}^{K+1}f_{W_{j}}(t)dt.
\end{align*}
Instead of directly computing the convolution of p.d.f.,  $F_{W_{\sum}}(x)$ can be alternatively obtained by
\begin{align*}
F_{W_{\sum}}(x) = & \mathcal{L}^{-1}\left\lbrace \frac{1}{s} \mathbb{E}\left[\exp\left(-s W_{\sum}\right)\right]\right\rbrace \left(x\right) \\
= & \mathcal{L}^{-1}\left\lbrace \frac{1}{s} \left(\mathcal{L}\left\lbrace f_{W_j}\right\rbrace \left(s\right) \right)^K\right\rbrace \left(x\right),
\end{align*}
where $\mathcal{L}$ and $\mathcal{L}^{-1}$ denote \emph{Laplace transform} and its inversion, respectively, and $s > 0$.

We are ready to derive the upper and lower bounds using $F_Y(*)$ and $F_{W_{\sum}}(*)$, shown in the following theorem.

\begin{theorem}\label{thm:performance bounds}
Let $F_{Y}\left(*\right)$ and $ F_{W_{\sum}}\left(*\right) $ denote the c.d.f. of $Y$ and $ W_{\sum} $, respectively, then we have
\begin{align*}
\underline{\mathcal{B}}\left(x\right)\leq \mathbb{P}\left(\frac{Y}{N_0 + W_{\sum}}\leq x\right) \leq  \overline{\mathcal{B}}\left(x\right),
\end{align*}
where $\underline{\mathcal{B}}\left(x\right) $ and  $ \overline{\mathcal{B}}\left(x\right)  $ are respectively given by
\begin{align*}
\underline{\mathcal{B}}\left(x\right)\triangleq
\sup_{t \geq 0} \left\lbrace F_{Y}\left((N_0 + t)x\right)- F_{W_{\sum}}\left(t\right)\right\rbrace,
\end{align*}
and
\begin{align*}
\overline{\mathcal{B}}\left(x\right)\triangleq 1+ 
\inf_{t \geq 0} \left\lbrace F_{Y}\left((N_0 + t)x\right)- F_{W_{\sum}}\left(t\right) \right\rbrace.
\end{align*}
\end{theorem}
\begin{IEEEproof}
For the upper bound, for any $t \geq 0$, we have
\begin{align*}
\mathbb{P}\left(\frac{Y}{N_0 + W_{\sum}}\leq x\right)
= & \mathbb{P}\left(\frac{Y}{N_0 + W_{\sum}}\leq x, W_{\sum} \geq t\right) \\
 & + \mathbb{P}\left(\frac{Y}{N_0 + W_{\sum}}\leq x, W_{\sum} \leq t\right)\\
\leq & \mathbb{P}\left(W_{\sum} \geq t\right)\! + \! \mathbb{P}\left(Y\leq \left( N_0 \! + \! t\right) x\right),
\end{align*}
then, the upper bound $\overline{\mathcal{B}}\left(x\right)$ can be immediately obtained.

For the lower bound, likewise,
\begin{align*}
\mathbb{P}\left(\frac{Y}{N_0 + W_{\sum}}\geq x\right)
= & \mathbb{P}\left(\frac{Y}{N_0 + W_{\sum}}\geq x, W_{\sum} \geq t\right) \\
 & + \mathbb{P}\left(\frac{Y}{N_0 + W_{\sum}}\geq x, W_{\sum} \leq t\right)\\
\leq & \mathbb{P}\left(W_{\sum} \leq t\right)\! + \! \mathbb{P}\left(Y\geq \left( N_0 \! + \! t\right) x\right)
\end{align*}
holds for any $t\geq 0$, and it subsequently gives
\begin{align*}
\mathbb{P}\left(\frac{Y}{N_0 + W_{\sum}}\leq x\right) \geq 
\mathbb{P}\left(Y\leq \left( N_0 \! + \! t\right) x\right) \! - \!  \mathbb{P}\left(W_{\sum} \leq t\right),
\end{align*}
which concludes the lower bound $\underline{\mathcal{B}}\left(x\right)$.
\end{IEEEproof}

Note that given $ \mathbf{Q}_1 $ and $ \varepsilon_1^r $,   the outage probability can be expressed as the c.d.f. of $ \gamma_{1} $, i.e.,
\[ p_{1,\mathrm{out}}  (R_{\mathrm{th}})\triangleq \mathbb{P} (R_1 < R_{\mathrm{th}} ) =  F_{\gamma_{1} }\left(2^{R_{\mathrm{th}}/W} -1 \right),\]
where  $R_{\mathrm{th}}$ denotes the rate threshold. Therefore tight bounds on the c.d.f. are essential in evaluating the performance.

\section{Performance Evaluation}\label{sec:performance evaluation}
\begin{figure}
\centering
\includegraphics[width=.7\columnwidth]{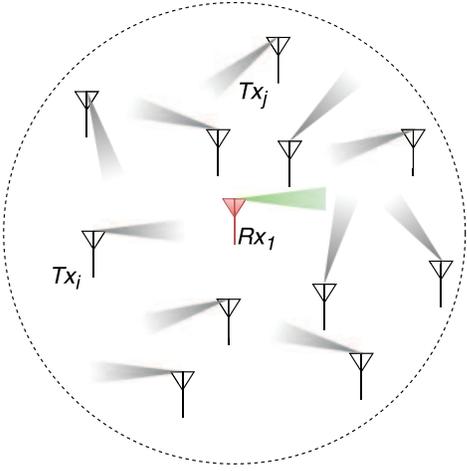}
\caption{Illustration of multiple interfering transmitters in the area of interest.}
\label{fig:example network}
\end{figure}
 
We consider a $ 60 $~GHz indoor wireless access network within a circular space of radius $ R_0 = 15 $ meters, as illustrated in Fig.~\ref{fig:example network}, where there are in total $ N=n$ concurrent transmissions.
The receiving node  in focus, $\mathrm{RX}_1$, is located at the center of a circular area and there are totally $K=n-1$ interfering transmitters distributed uniformly at random within the area of interest, randomly oriented in  a uniform manner. Results by numerical and Monte-Carlo methods are presented to investigate the accuracy of the bounds, and the sensitivity of outage probability and average throughput against beam patterns and misalignment.   To simplify the performance evaluation, all nodes are assumed to be placed on the same horizontal plane. 
The common system parameters are summarized in Table~\ref{tab:system parameter} and the p.d.f. of link lengths can be found in~\cite{yang2015maximum}.
Our evaluation consists of the  two parts:
\begin{enumerate}
\item Numerical results to validate the bounds for the fixed typical receiver  $\mathrm{RX}_1$, as depicted in Fig.~\ref{fig:example network}.
\item Simulation results to evaluate the average performance of randomly deployed typical receivers. 
\end{enumerate}

\begin{table}
\small
\renewcommand{\arraystretch}{1.5}
\centering
\caption{System Parameters}
\begin{tabular}{|c|c|c|}
\hline
\textbf{Parameter} & \textbf{Notation} & \textbf{Value} \\
\hline
\hline
Wavelength & $ \lambda $ & $ 5\times 10^{-3} $ m\\
\hline
Bandwidth & $ W $ & $ 500 $ MHz\\
\hline
Transmit Power & $ P_{t} $ & $ 1 $ mW\\
\hline
Main-lobe Beamwidth & $\theta_m$ & $ \left[\frac{\pi}{12},\frac{\pi}{2}\right] $ \\
\hline
3dB-Beamwidth Ratio & $\eta$ & $ \left(0,1\right) $ \\
\hline
Misalignment Deviation & $\rho$ & $ \left[0,\frac{1}{6}\right] $ \\
\hline
Noise Power Density & $N_0/W$ & $ -114 $ dBm/MHz\\
\hline
Radius of Circular Hall & $ R_0 $ & $ 15 $ m\\
\hline
Path Loss Exponent & $\alpha$ & $ 2.45 $ \\
\hline
Link Numbers & $ n $ & $ \leq 30 $ \\
\hline
\end{tabular}\label{tab:system parameter}
\end{table}

\subsection{Bounds for Fixed Typical Receiver at The Center}

\begin{figure}
\centering
\includegraphics[width=.85\columnwidth]{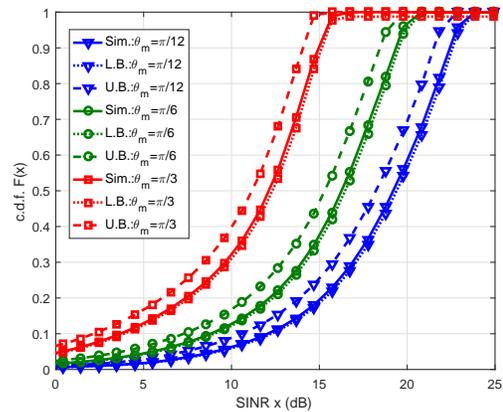}
\caption{The lower and upper bounds of c.d.f.s for $ \gamma_{1} $ and simulated results, with respect to diverse $ \theta_{m} $, where  $ n=11 $, $ \eta=0.4 $, and $ \rho=\frac{1}{20}$.}
\label{fig:cdf of bounds vs. bw}
\end{figure}

We validate the bounds derived in Theorem~\ref{thm:performance bounds} for the fixed typical receiver and investigate  the impact of  the main-lobe beamwidth $ \theta_m $ and the half-power beamwidth ratio $ \eta $ on the c.d.f. of SINR. 
The lower and upper bounds on the c.d.f. of SINR  are illustrated in Fig.~\ref{fig:cdf of bounds vs. bw}, where $ n=11 $, $ \eta = 0.4 $, and $ \rho=\frac{1}{20} $. To investigate the impact of $ \theta_m $ and the associated factor $ \eta $, we consider the following three distinct values of beamwidth, i.e., $ \theta_m = \frac{\pi}{12} $, $ \frac{\pi}{6} $, and $ \frac{\pi}{3} $, respectively. In general, the derived bounds in Theorem~\ref{thm:performance bounds} behave well for all groups, which validates the feasibility of applying our upper and lower bounds in analyzing the actual system performance. Note that in all combinations we have considered here, the lower bound outperforms its upper counterpart. Furthermore,  considerable performance gain can be achieved by narrowing down the beamwidth. For instance, when $ \theta_m =\frac{\pi}{3} $ reduces to its half, i.e., $ \theta_m=\frac{\pi}{6} $, there is roughly  $ 4$~dB gain,  and there is another $ 3 $~dB gain when $ \theta_m $ keeps going down to $ \theta_m=\frac{\pi}{12} $.

\begin{figure}
\centering
\includegraphics[width=.9\columnwidth]{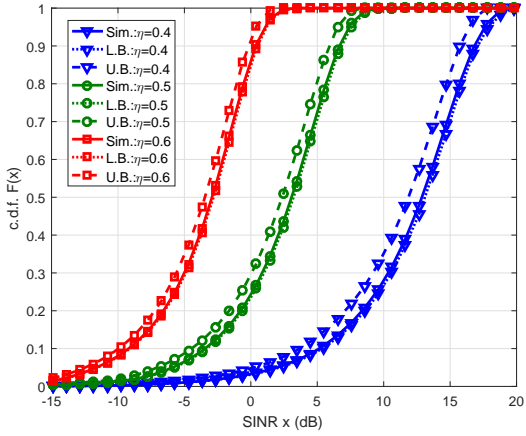}
\caption{The lower and upper bounds of c.d.f.s for $ \gamma_{1} $ and simulated results, with respect to diverse $ \eta $, where $ n=21 $, $ \theta_m = \frac{\pi}{6} $ and $ \rho=\frac{1}{20} $.}
\label{fig:cdf of bounds vs. eta}
\end{figure}
In Fig.~\ref{fig:cdf of bounds vs. eta}, we demonstrate the bound performance against the factor $ \eta $, where $ n=21 $, $ \theta_m = \frac{\pi}{6} $ and $ \rho=\frac{1}{20} $. Again, both the upper and lower bounds are very tight. 
Despite of a narrow beamwidth, i.e., $ \theta_m = \frac{\pi}{6} $, is employed, there is still huge performance difference for different $ \eta $. 
As shown in  the figure, a substantial gain can be achieved by decreasing $ \eta $ (i.e., a faster attenuating main-lobe).
For instance, the performance gains roughly $ 6 $~dB when decreasing $ \eta $ from $ 0.6 $ to $ 0.5 $, while roughly $ 10 $~dB gain can be achieved by $ \eta=0.4 $. This indicates the great importance of $ \eta $ in the antenna design.

The above results show that, both the main-lobe beamwidth and the half-power beamwidth ratio are crucial factors that determine the performance. In what follows, we will consider the scenario where the typical receiver is randomly located.
 
\subsection{Simulations for Randomly Located Typical Receiver}
In contrast to the aforementioned scenario with fixed $ \mathbf{Q}_1 $ and $ \varepsilon_1^{r} $, we here focus on the situation  where the typical receiver is randomly located, and the misalignment is not given. Besides the outage probability, in this section, the  sum throughput and the average throughput are also evaluated.  

\begin{figure}
\centering
\includegraphics[width=.9\columnwidth]{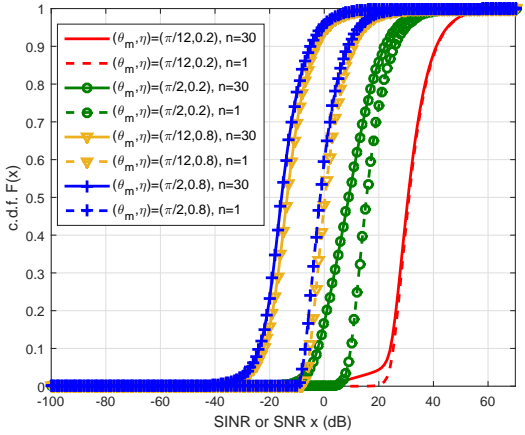}
\caption{The c.d.f.s of SINR $ \gamma_{1} $ with diverse $ (\theta_{m}, \eta) $ pairs for $ n=1 $ or $ 30 $, where $ \rho=\frac{1}{20}$.}
\label{fig:cdf of sinr and snr}
\end{figure}

In Fig.~\ref{fig:cdf of sinr and snr},  we illustrate the c.d.f. of $\gamma_{1}$ with the fixed misalignment derivation to main-lobe beamwidth ratio $\rho{=}\frac{1}{20}$, where the main-lobe beamwidth $\theta_{m}$ is set to be $ \frac{\pi}{12} $ or $ \frac{\pi}{2} $, and the half-power beamwidth ratio  $\eta$ is chosen as $0.2$ or $0.8$. We denote by $ F_{{\gamma}_{1}}^{(1)}\left(x\right)$ and $ F_{{\gamma}_{1}}^{(30)}\left(x\right)$ the outage probabilities associated with $ n=1 $ (hence no concurrent transmission interference) and $ n=30 $, respectively.
%The c.d.f. upper bound $ F_{\overline{\gamma}_{1}}\left(x\right) $ is obtained when there is only one single pair of active nodes (hence no concurrent transmission interference) and the lower bound  $ F_{\underline{\gamma}_{1}}\left(x\right)$ is obtained when all the $ n=30 $ pairs are active.
For  $ \left(\theta_{m},\eta\right){=}\left(\frac{\pi}{12},0.2\right)$ that corresponds to the scenario where the main-lobe is narrow and the beam attenuates fast,  there is only a small gap between $ F_{\gamma_1}^{(30)}\left(x\right) $ and $ F_{\gamma_1}^{(1)}\left(x\right) $. This is in line with our intuition that when the receive beam is very narrow and the beam misalignment is small, the degradation caused by concurrent transmission interference is not significant except for users with low signal power. If we then hold $ \eta{=}0.2$ (i.e., fast attenuation beam) but increase the main-lobe beamwidth $ \theta_{m}$ from $\frac{\pi}{12}$ to $ \frac{\pi}{2}$, the gap between $ F_{\gamma_1}^{(30)}\left(x\right) $ and $ F_{\gamma_1}^{(1)}\left(x\right) $, with a gap of around $5$dB at $10$-percentile and 1dB at $90$-percentile. Interestingly, if we set $\eta{=}0.8$, i.e., the main-lobe has almost a constant-gain top, the gap between upper and lower bounds remains almost a constant of 8dB from $10$-percentile up to $90$-percentile, and the influence of the main-lobe beamwidth $ \theta_{m}$ is very limited: less than 1~dB gap between $\theta_{m}=\frac{\pi}{12}$ and $\theta_m=\frac{\pi}{2}$. Therefore, when beam misalignment is small, the main-lobe attenuation speed, quantified by $\eta$, dominates the sensitivity to concurrent transmission interference.

\begin{figure}
\centering
\includegraphics[width=.9\columnwidth]{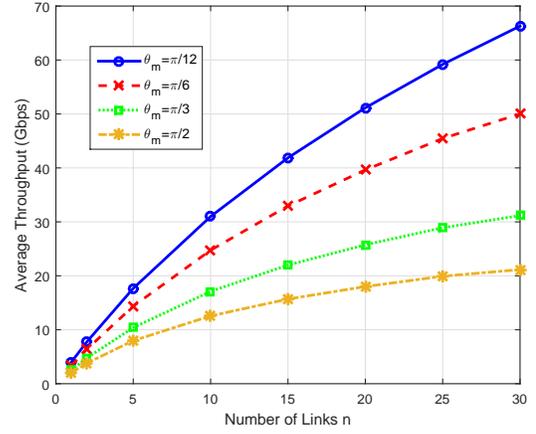}
\caption{The average sum throughput with varying $  n \leq 30 $, where $ \theta_{m} = \frac{\pi}{12}, \frac{\pi}{6}, \frac{\pi}{3} $ or $ \frac{\pi}{2} $,  $ \rho{=}\frac{1}{20}$, and $\eta {=} \frac{1}{2.6}$. }
\label{fig:overall throughput}
\end{figure}

In Fig.~\ref{fig:overall throughput} we plot the average sum throughput as a function of the number of active links $ n $ ranging from $ 1 $ to $ 30 $,
where the main-lobe beamwidth $ \theta_{m} $ is set to $ \frac{\pi}{12} $, $ \frac{\pi}{6} $, $ \frac{\pi}{3} $, and $ \frac{\pi}{2}$, respectively, with fixed misalignment derivation to main-lobe beamwidth ratio $ \rho{=}\frac{1}{20}$ and half-power beamwidth ratio $\eta {=} \frac{1}{2.6}$, which is adopted from the experiment validation in~\cite{toyoda2006reference}.
As the number of active links increases, the average sum throughput increases much faster for narrow beam $\theta_m{=}\frac{\pi}{12}$ compared to wide beam $\theta_m{=}\frac{\pi}{2}$, as determined by the slopes of the curves. This is in line with our observations from  Fig.~\ref{fig:cdf of sinr and snr} where, when the main-lobe attenuates fast, the links with small main-lobe beamwidth are more or less noise/power limited whereas the links with large main-lobe are interference limited. 

\begin{figure}
\centering
\includegraphics[width=.9\columnwidth]{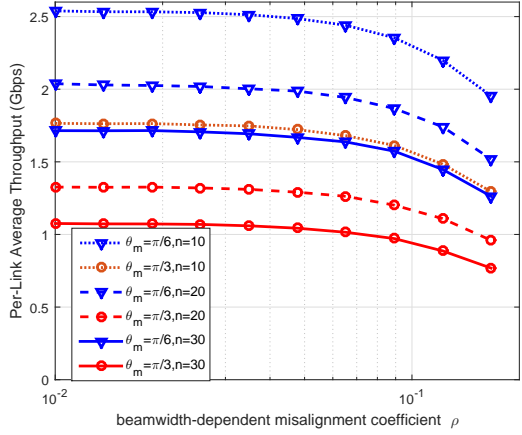}
\caption{Sensitivity of the per-link average throughput against $\rho$, where $\eta=\frac{1}{2.6}$, $ \theta_{m} = \frac{\pi}{6} $ or $ \frac{\pi}{3} $, and $n=10, 20$ or $30$.}
\label{fig:rho on average throughput}
\end{figure}

In Fig.~\ref{fig:rho on average throughput} we investigate the sensitivity of the per-link average throughput against $ \rho $ with a fixed half-power beamwidth ratio $\eta=\frac{1}{2.6}$. We investigate two groups with $ \theta_m=\frac{\pi}{6} $ and $ \frac{\pi}{3} $, respectively, with the number of active links $ n=10 $, $ 20 $ or $ 30 $.
For any given $\rho$, the per-link average throughput will decrease significantly as the main-lobe beamwidth $\theta_m$ and/or the number of active links $n$ increases, which clearly attributes to the increase of the concurrent transmission interference.
Such per-link performance degradation (gap among different lines) decreases slightly as the misalignment increases.
%Another observation is that although the per-link throughput degradation decreases as the number of links $n$
For fixed $n$ and $\theta_m$, the per-link average throughput remains stable for  $ \rho<0.05$ and the degradation grows up to about $30\%$ as  $\rho \to \frac{1}{6} $. 
%This provides a more comprehensive image of the performance sensitivity against $\rho$ shown in Fig.~\ref{fig:rho on outage}.
Regarding the practical significance, on the one hand, it is beneficial to reduce beam misalignment, but the reward is diminishing as $ \rho$ becomes smaller. Since a high alignment precision  indicates a high overhead/cost in practical implementations, a quantitative evaluation of the performance loss is crucial to seek the proper  trade-off between the performance and cost.
On the other hand, the performance degradation caused by  $\rho$ remains almost the same as the main-lobe beamwidth increases from $\frac{\pi}{6}$ to $\frac{\pi}{2}$, which clearly justifies our effort in quantifying the misalignment via $\rho$.

\section{Conclusions}\label{sec:conclusion}
We study the impact of antenna beam misalignment and beam patterns on the system performance of $ 60 $ GHz wireless access. A practical directional antenna model that considering both the main-lobe and side-lobe gains is applied. We  introduced two main-lobe beamwidth-dependent parameters, namely, the half-power beamwidth ratio $\eta$ to quantify the main-lobe attenuation speed, and the misalignment deviation ratio $\rho$ to quantify the concentration of beam misalignment. We derived the probability distribution of the SINR, and  developed tight upper and lower bounds to facilitate tractable performance analysis. Our numerical results demonstrate the tightness of our derived upper and lower bounds, and reveal that the parameter $ \eta $  plays a critical role in enhancing the network performance. Furthermore, we quantified the sensitivity of performance deterioration with respect to beam misalignment and aggregated interferences from neighboring concurrent transmissions. 
Our results reveal the importance of the two key parameters $\eta$ and $\rho$ in system design to balance the impact of beam misalignment and concurrent transmission interference.

\appendix
\section{Product of Two Random Variables}\label{appendix:1}
Without loss of generality, assuming $ c $ is a non-zero constant scalar, let $ X = cYZ $ be a function of two positive random variables $ Y $ and $ Z $, with marginal p.d.f.s $ f_{Y}(y) $ and $ f_{Z}(z) $, accordingly, where $ c $. Introducing an auxiliary random variable $ v = z $, with $ x=cyz $, we can obtain $ y=\frac{x}{cv} $ and $ z=v $, respectively. Through the function of multivariate random variables \cite{papoulis2002probability}, we have
\begin{equation*}
f_{X,V}(x,v)=\frac{f_{Y,Z}(y,z)}{|\mathcal{J}_{x,v}\left(y,z\right)|},
\end{equation*}
where $ \mathcal{J}_{x,v}\left(y,z\right) $ is given by
\begin{equation*}
\mathcal{J}_{x,v}\left(y,z\right)=
\begin{vmatrix}
\frac{\partial x}{\partial y} & \frac{\partial x}{\partial z} \\ 
\frac{\partial v}{\partial y} & \frac{\partial v}{\partial z} 
\end{vmatrix} 
=
\begin{vmatrix}
cz & cy\\ 
0 & 1 
\end{vmatrix}
=cv,
\end{equation*}
thus we have
\begin{equation*}
f_{X,V}(x,v)=\left(|c|v\right)^{-1}f_{Y,Z}\left(\frac{x}{cv},v\right).
\end{equation*}

Finally, the p.d.f. $ f_{X}(x) $ can be immediately obtained by the integral over all possible $ v $. That is,
\begin{equation*}
f_{X}(x)=\int\limits_{v\in \mathcal{S}_{z}}\left(|c|v\right)^{-1}f_{Y,Z}\left(\frac{x}{cv},v\right)dv,
\end{equation*}
where $ \mathcal{S}_{z} $ corresponds to the domain of marginal p.d.f. of $ Z $, namely, $ f_{Z}(z) $.

Particularly, if $ Y $ and $ Z $ are independent random variables, we further have
\begin{equation*}
f_{X}(x)=\int\limits_{v\in \mathcal{S}_{z}}\left(|c|v\right)^{-1}f_{Y}\left(\frac{x}{cv}\right)f_{Z}\left(v\right)dv.
\end{equation*}

\end{document}